\documentclass[preprintnumbers,aps,preprint,nofootinbib]{revtex4-1}

\usepackage{booktabs}
\usepackage{footnote}

\usepackage{lipsum}
\usepackage{amsmath}    
\usepackage{amssymb}    
\usepackage{color}         
\usepackage{graphicx}     
\usepackage[T1]{fontenc}
\usepackage{times}         
\usepackage{mathrsfs} 
\usepackage{myterms}
\usepackage{hyperref} 
\hypersetup{
colorlinks=true,       
linkcolor=blue,        
citecolor=blue,        
filecolor=magenta,     
urlcolor=blue          
}
\usepackage[all]{hypcap} 

\usepackage[none]{hyphenat} 

\usepackage{color,xcolor,fancybox,epsf,rotating,colordvi}

\begin{document}

\preprint{\parbox{1.6in}{\noindent ~~~~~~}}

\title{Light dark matter confronted with the 95 GeV diphoton excess}

\author{Weichao Li}
\affiliation{School of Physics and Technology, Wuhan University, Wuhan 430072, China}

\author{Haoxue Qiao}
\affiliation{School of Physics and Technology, Wuhan University, Wuhan 430072, China}

\author{Kun Wang}
\affiliation{College of Science, University of Shanghai for Science and Technology, Shanghai 200093, China}

\author{Jingya Zhu}
\email[]{zhujy@henu.edu.cn} 
\affiliation{School of Physics and Electronics, Henan University, Kaifeng 475004, China}

\date{\today}

\begin{abstract}
The correlation between Higgs-like scalars and light dark matter is an interesting topic, especially now that a $125\GeV$ Higgs was discovered and dark matter (DM) searches got negative results. 
The $95\GeV$ excess reported by the CMS collaboration with $132\fbm$ data recently, and the DM search results by XENONnT and LZ collaborations motivate us to revise that. 
In this work, we study that in the GUT-scale constrained (GUTc) Next-to-Minimal Supersymmetric Model (NMSSM), where most parameters are input at the GUT scale, but with scalar and gaugino masses not unified there. 
In the calculation we also consider other recent experimental constraints, such as Higgs data, Supersymmetry (SUSY) searches, DM relic density, etc. 
After detailed analysis and discussion, we find that: 
(i) The light DM can be bino- or singlino-dominated, but can be mixed with minor components of Higgsino. 
(ii) Both cases can get right relic density and sizable Higgs invisible decay, by adjusting the dimensionless parameters $\lambda, \kappa$, or suitably mixing with Higgsino. 
(iii) Both cases can have four funnel annihilation mechanisms, i.e., annihilating through $Z, a_1, h_2, h_1$. 
(iv) Samples with right relic density usually get weak signal of Higgs invisible decay at future lepton collider, but the  $95\GeV$ scalar can have sizable $b\bar{b}$ signal. 
\end{abstract}


\maketitle

\section{Introduction}
\label{intro}
The Standard Model (SM) became a successful theory, especially after the Higgs boson was discovered in 2012 \cite{ATLAS:2012yve, CMS:2012qbp}, and later proved to be very SM-like \cite{ATLAS:2022vkf, CMS:2022dwd, Salam:2022izo}. 
However, whether there is an additional Higgs-like scalar is a natural and still open question. 
About twenty years ago, a $Zb\bar{b}$ anomaly was reported with an invariant mass of $b\bar{b}$ at about $98\GeV$ \cite{LEPWorkingGroupforHiggsbosonsearches:2003ing}. 
In 2018, the CMS collaboration reported a $95.3\GeV$ diphoton anomaly with about $56\fbm$ data \cite{CMS:2018cyk}.
In 2022, CMS also reported a $\tau^+\tau^-$ anomaly at $95\sim 100\GeV$ \cite{CMS:2022goy}. 
Recently in March 2023, CMS updated their results of low mass diphoton measurement with $132\fbm$ data, and the excess still stands there at about $95.4\GeV$ \cite{CMS:2023yay}. 
Considering the nearness in the mass region, it is natural to interpret these anomalies by one additional scalar of about $95\sim 100\GeV$ \cite{Ahriche:2023hho, Cao:2023gkc, Ellwanger:2023zjc, Maniatis:2023aww, Azevedo:2023zkg, Chen:2023bqr, Ahriche:2023wkj, Arcadi:2023smv, Borah:2023hqw, Dutta:2023cig, Aguilar-Saavedra:2023tql, Ashanujjaman:2023etj, Belyaev:2023xnv, Escribano:2023hxj, Azevedo:2023zkg, Biekotter:2023jld}. 

Besides, the SM also meets with other challenges, such as naturalness, grand unification, dark matter, etc. 
The gravitational effects of dark matter have been known for nearly a century, and it is widely believed that it takes part in weak interaction. 
So, around the world, tens of experiments attempt to detect its weak effect, i.e., scatter with a nucleus. 
Recently the LZ \cite{LZ:2022lsv} and XENONnT \cite{XENON:2023cxc} collaborations updated their results of direct searches, suggesting stronger constraints to new physics models with dark matter. 
Considering the negative results of direct DM searches and relatively successful ones of Higgs measurement, it is interesting to survey the correlations between light dark matter and Higgs-like scalars. 
After Higgs was discovered, Higgs decay to DM and DM funnel annihilate through Higgs boson has been studied in several works \cite{Wang:2020dtb, Cao:2012im, Zhao:2020trt, Cho:2020ftg, Wang:2016lvj, Tang:2015uha, Han:2014nba, Barman:2020vzm, Guchait:2020wqn}. 
It is also interesting to survey the implications of $95\GeV$ diphoton anomaly on light DM. 

Supersymmetry (SUSY), by introducing a new boson-fermion symmetry, can gracefully solve most problems of the SM, and is widely considered as a candidate theory of new physics beyond the SM \cite{Martin:1997ns}. 
The Next-to-Minimal Supersymmetric Model (NMSSM) has in addition two Higgs bosons and one neutralino to the minimal one (MSSM), thus has more abundant phenomenology on Higgs, dark matter, and SUSY searches \cite{Gunion:2005rw, Belanger:2005kh, Ellwanger:2009dp, Maniatis:2009re, Ferrara:2010in, Ellwanger:2011aa,  King:2012is, Cao:2012fz, Borah:2023zsb, Jia:2023xpx, Binjonaid:2023rtc, Dao:2023kzz, Cao:2023juc, Bisal:2023mgz, Wang:2019biy, Wang:2020tap, Ma:2020mjz, Wang:2021fsz, Wang:2021lwi, Li:2022etb}. 
In this work, we study light DM confronted with the $95\GeV$ diphoton anomaly in the GUT-scale constrained (GUTc) Next-to-Minimal Supersymmetric Model (NMSSM), where most parameters are input at the GUT scale, but with Higgs and gaugino masses, respectively, not unified there. 
Besides constraints on DM and Higgs, we also consider other related constraints, such as SUSY searches, B physics, muon $g-2$ \cite{Muong-2:2021ojo, Muong-2:2023cdq}, etc. 

The other parts of this paper are arranged below. 
In Sec. \ref{ana}, we succinctly present an introduction to NMSSM, mainly in the Higgs and neutralino sectors with analytical equations.
Then in Sec. \ref{num}, we present numerical calculations and discussions. 
Finally, we draw our conclusions in Sec. \ref{conc}.

\section{The Higgs and neutralino sectors in NMSSM}
\label{ana}

NMSSM introduce a singlet superfield $\hat{S}$ to MSSM, with superpotential $W^{\mathrm{NMSSM}}$ under $\mathbb{Z}_3$ symmetry related to that of MSSM $W^{\mathrm{MSSM}}$ by: 
\begin{equation}
	W^{\mathrm{NMSSM}}=W^{\mathrm{MSSM}}_{\mu\to \lambda \hat{S}} + \frac{\kappa \hat{S}^3}{3} \,,
\end{equation}
where $\lambda, \kappa$ are two dimensionless couplings. 
Thus when the singlet acquires Vacuum Expectation Value (VEV) $v_S$, NMSSM generates the $\mu$ term naturally through $\mu_{\mathrm{eff}} = \lambda v_S$. 

In the Higgs sector with CP conserving, when electroweak symmetry breaks, i.e., $\rm SU(2)_{L}\otimes U(1)_Y \to \rm U(1)_{EM}$, the doublet and singlet scalars with the same residual quantum number mix to form three CP-even and two CP-odd mass-eigenstate Higgs bosons, i.e., $h_{1,2,3}$ and $a_{1,2}$ in increasing mass order, respectively. 
The coefficients of doublet $H_{u,d}$ and singlet $S$ in $h_{1,2,3}$ and $a_{1,2}$ can be denoted as $S_{i\alpha}$ and $P_{j\alpha}$, with $i,\alpha=1,2,3$ and $j=1,2$, respectively. 

In the neutralino sector, the neutral SUSY partners bino $\tilde b$, wino $\tilde W$, Higgsino $\tilde H_{u,d}$, and singlino $\tilde{S}$, with the same residual quantum number, can mix to form five neutralinos $\tilde \chi_i$ in increasing mass order, with coefficients of $N_{ij}$ ($i,j=1,...,5$) in each neutralino. 
In this work, the first neutralino $\tilde\chi^0_1$ is required as the lightest SUSY particle (LSP), and thus DM candidate for $R$ parity conserved. 

The couplings between Higgs bosons and neutralinos can be writen as \cite{Ellwanger:2004xm}: 
\begin{eqnarray}
	C_{h_\alpha \tilde\chi^0_i \tilde\chi^0_j} 
	&=& 
	- \frac{g_1}{2} (S_{\alpha 1} \Pi_{ij}^{13} - S_{\alpha 2} \Pi_{ij}^{14}) 
	+ \frac{g_2}{2} (S_{\alpha 1} \Pi_{ij}^{23} - S_{\alpha 2} \Pi_{ij}^{24})
	\nonumber \\
	&& +\frac{\lambda}{\sqrt{2}} (S_{\alpha 1} \Pi_{ij}^{45} 
	+ S_{\alpha 2} \Pi_{ij}^{35} + S_{\alpha 3} \Pi_{ij}^{34}) 
	\nonumber \\
	&& -\sqrt{2} \kappa S_{\alpha 3} N_{i5} N_{j5} 
	\\ 
	C_{a_\beta \chi^0_i \chi^0_j} 
	&=& i\left[\frac{g_1}{2} (P_{\beta 1} \Pi_{ij}^{13} 
	- P_{\beta 2} \Pi_{ij}^{14}) 
	- \frac{g_2}{2} (P_{\beta 1} \Pi_{ij}^{23} 
	- P_{\beta 2} \Pi_{ij}^{24}) \right. 
	\nonumber \\ 
	&& ~~~+\frac{\lambda}{\sqrt{2}} (P_{\beta 1} \Pi_{ij}^{45} + P_{\beta 2} \Pi_{ij}^{35} + P_{\beta 3} \Pi_{ij}^{34})
	\nonumber \\  
	&& \left. ~~~ - \sqrt{2} \kappa P_{\beta 3} N_{i5} N_{j5} \right] \,,
\end{eqnarray}
where $\Pi^{kl}_{ij} = N_{ik}N_{jl}+N_{il}N_{jk}$. 

Considering $h_1$ as a lighter Higgs to interpret the $95\GeV$ excesses, it should be singlet-dominated, and also the CP-odd one $a_1$. 
When the LSP $\tilde\chi^0_1$ is bino-dominated, the couplings between light Higgs and LSP $\tilde\chi^0_1$ can be approximatively written as: 
\begin{eqnarray}\label{h1bobo}
	C_{h_1 \tilde\chi^0_1 \tilde\chi^0_1} 
	&\approx& 
	g_1 N_{11} (S_{11} N_{14} - S_{12} N_{13}) 
	\nonumber \\
	&& + \sqrt{2}S_{13} (\lambda N_{13} N_{14} 
	- 2\kappa N_{15} N_{15})
	\\ 
	C_{a_1 \tilde\chi^0_1 \tilde\chi^0_1} 
	&\approx& i g_1 N_{11} (P_{11} N_{13} 
	- P_{12} N_{14}) 
	\nonumber \\
	&& + \sqrt{2} i P_{13} (\lambda N_{13}N_{14}
	-2 \kappa N_{15}N_{15}) \,, \label{a1bobo}
\end{eqnarray}
where $g_1$ is the $\rm U(1)_Y$ gauge coupling. 
While when the LSP $\tilde\chi^0_1$ is singlino-dominated, the couplings can be approximatively by: 
\begin{eqnarray}
	C_{h_1 \tilde\chi^0_1 \tilde\chi^0_1} 
	&\approx& 
	\sqrt{2} \left[ \lambda N_{15} (S_{11} N_{14} + S_{12} N_{13}) \right.
	\nonumber \\
	&& \left. ~~~~~~~~+S_{13} (\lambda N_{13}N_{14}-\kappa N_{15} N_{15}) \right]
	\label{h1soso} \\ 
	C_{a_1 \chi^0_1 \chi^0_1} 
	&\approx& 
	i\sqrt{2} \left[ \lambda N_{15} (P_{11} N_{14} + P_{12} N_{13})  \right.
	\nonumber \\ 
	&& \left. ~~~~~~~~ + P_{13} (\lambda N_{13}N_{14} - \kappa N_{15} N_{15}) \right] \,, \label{a1soso}
\end{eqnarray}

\section{Numerical results and discussions}
\label{num}
As a successive work, we first update our scan result in our former work \cite{Li:2022etb} with {\sf NMSSMTools -6.0.2}  \cite{Ellwanger:2006rn, Ellwanger:2005dv, Ellwanger:2004xm} and {\sf SModelS-v2.3} \cite{MahdiAltakach:2023bdn, Alguero:2021dig, Ambrogi:2017neo, Ambrogi:2018ujg, Dutta:2018ioj, Kraml:2013mwa} there. 
Then to focus on the correlation between light DM and $95\GeV$ excess, we add more samples with LSP mass between $46$ and $50\GeV$, i.e., about half of $h_1$ mass. 
We also introduce newly released experimental data, e.g., low-mass diphoton excess 2023 by the CMS collaboration \cite{CMS:2023yay}, and direct searches for DM by the LZ \cite{LZ:2022lsv} and XENONnT \cite{XENON:2023cxc} collaborations. 
Thus the surviving samples used here satisfy constraints of Higgs data, muon g-2, B physics, SUSY searches, DM density and direct searches, etc. 

\begin{figure*}[htbp]
	\centering
	\includegraphics[width=0.96\textwidth]{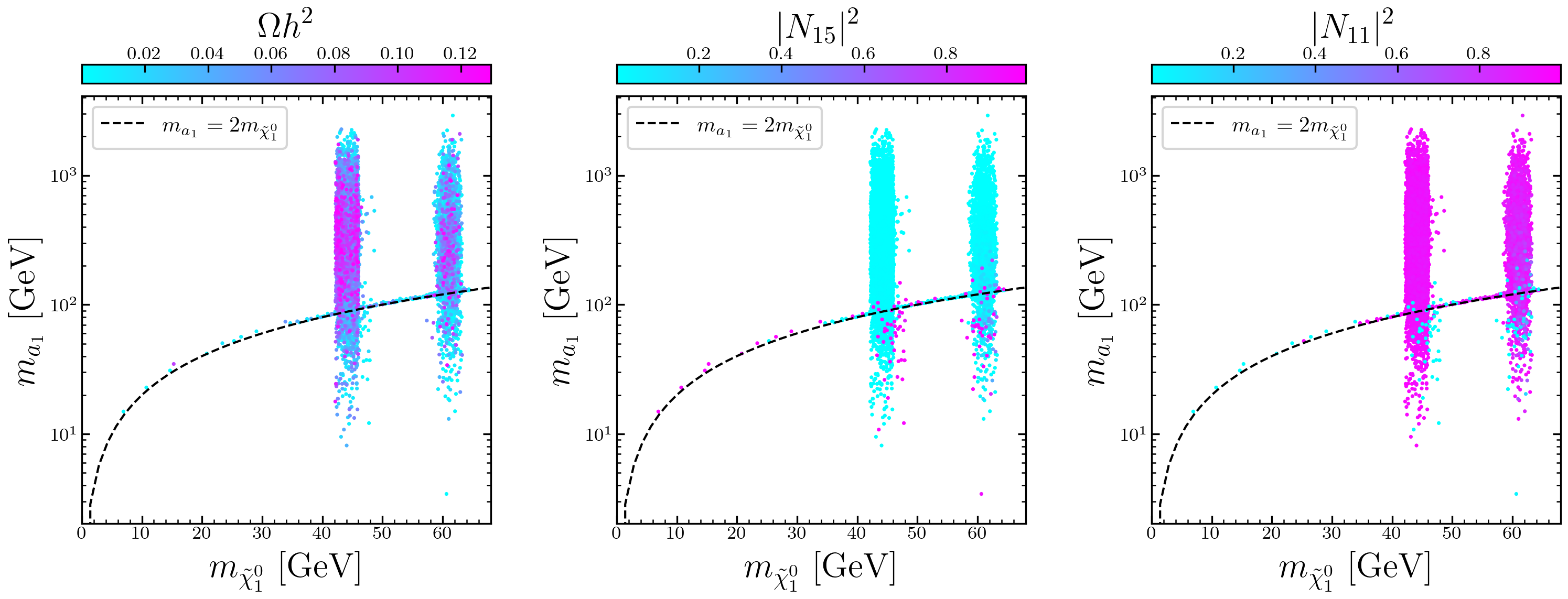}\\
	\includegraphics[width=0.96\textwidth]{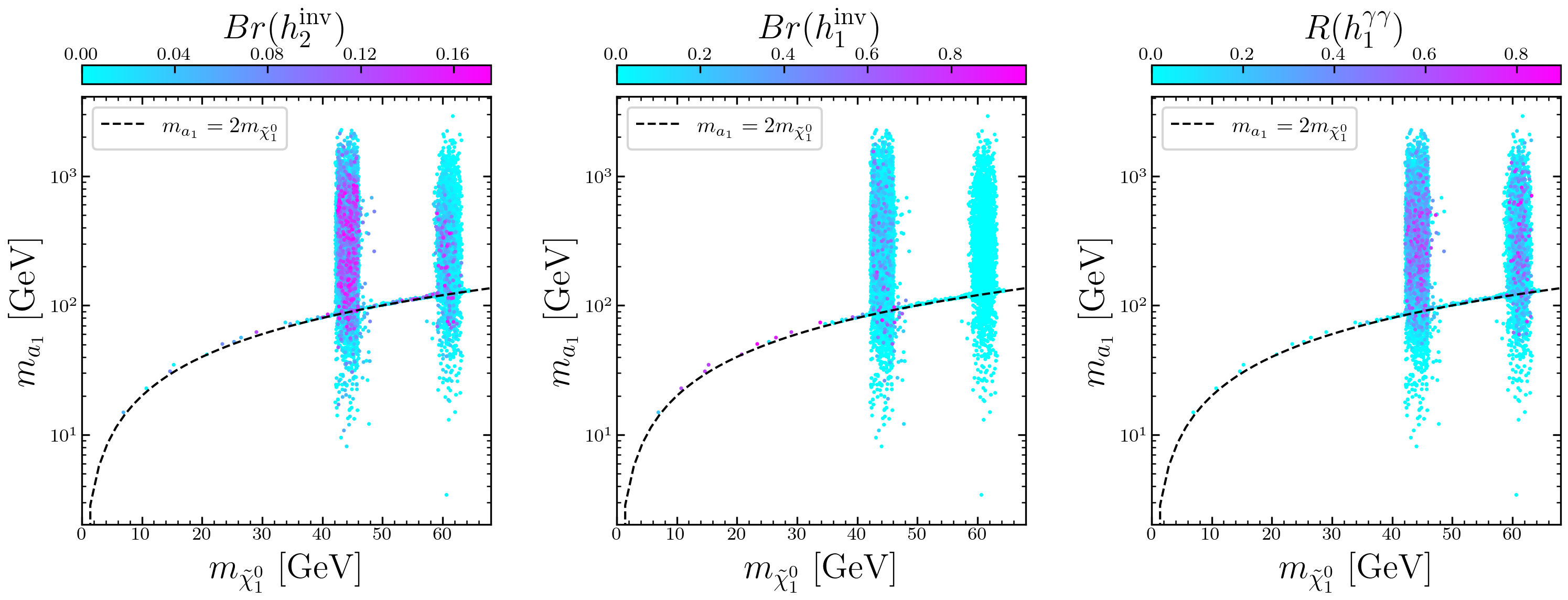}
	\vspace{-0.2cm}
	\caption{Surviving samples in the light pesudoscalar mass $m_{a_1}$ versus LSP mass $m_{\tilde\chi^0_1}$ planes, with colors indicating DM relic density $\Omega h^2$ (upper left), LSP singlino component $|N_{15}|^2$ (upper middle) and LSP bino component $|N_{11}|^2$ (upper right), $h_2$ invisible branching ratio $Br(h_2^{\rm inv})$ (lower left), $h_1$ invisible branching ratio $Br(h_1^{\rm inv})$ (lower middle), and $h_1$ diphoton signal rate $R(h_1^{\gamma\gamma})$ (lower right), respectively.}
	\label{fig1}
\end{figure*}

\begin{figure*}[htbp]
	\centering
	\includegraphics[width=0.96\textwidth]{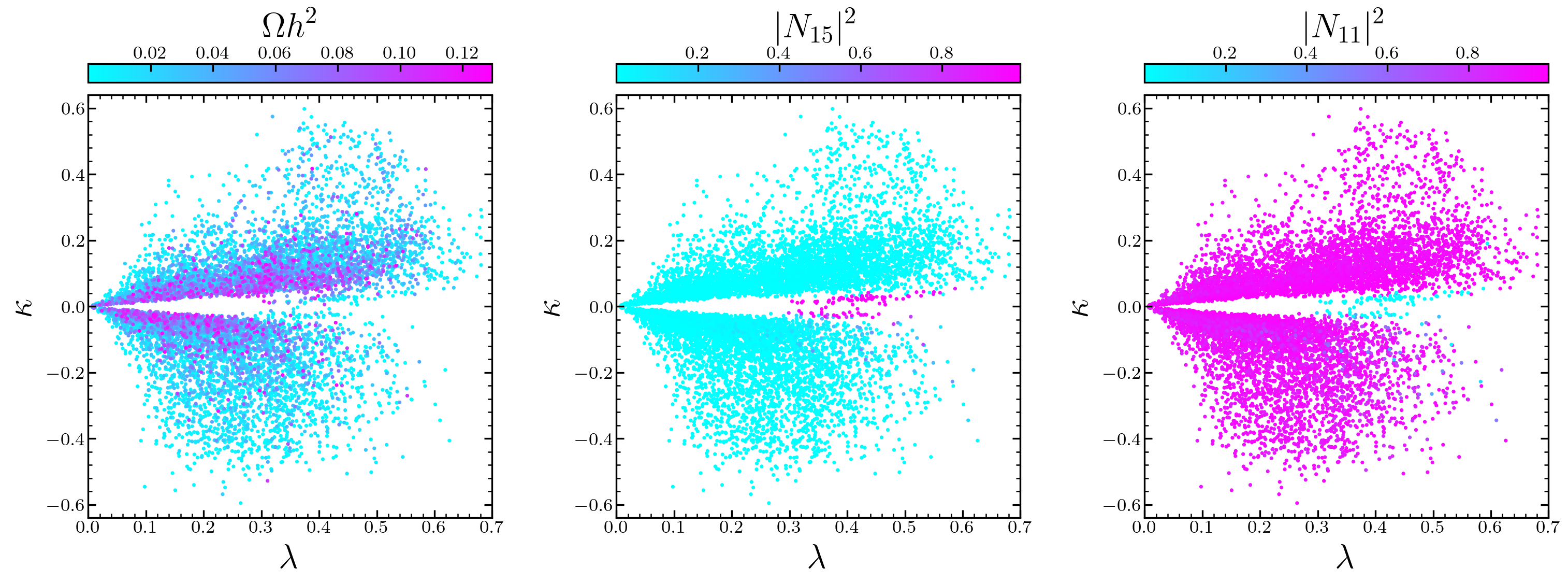}\\
	\includegraphics[width=0.96\textwidth]{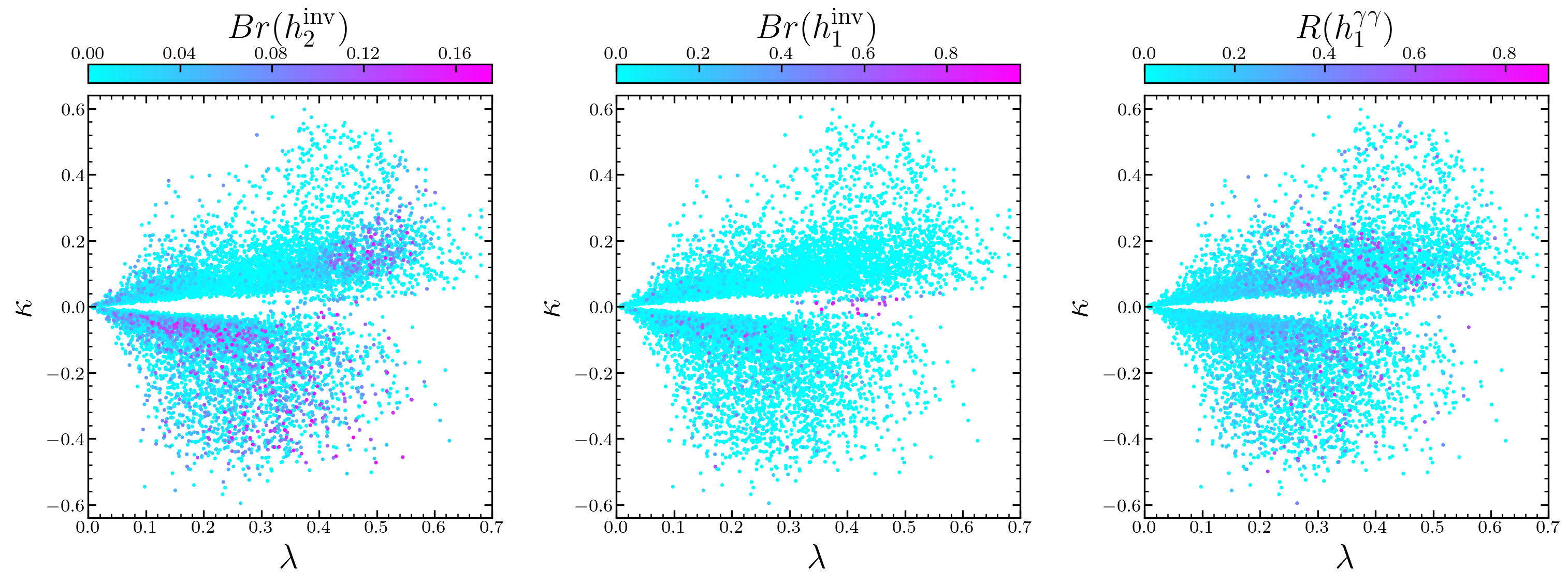}
	\vspace{-0.2cm}
	\caption{Same as in \ref{fig1}, but surviving samples in the $\kappa$ versus $\lambda$ planes.}
	\label{fig2}
\end{figure*}

\begin{figure*}[htbp]
	\centering
	\includegraphics[width=0.96\textwidth]{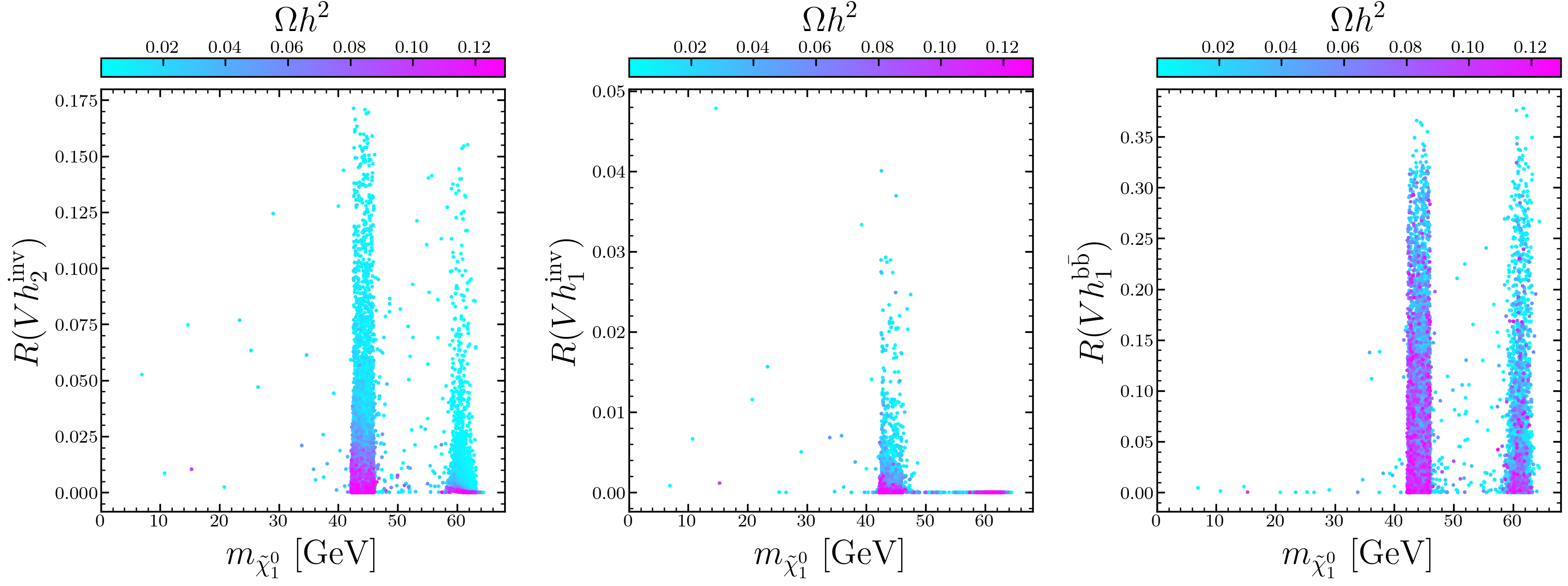}
	\vspace{-0.2cm}
	\caption{Surviving samples in the signal rates $R(V h_2^{\rm inv})$ (left), $R(V h_1^{\rm inv})$ (middle), and $R(V h_1^{\rm b\bar{b}})$ (right) versus LSP mass $m_{\tilde\chi^0_1}$ planes, respectively, with colors indicating DM relic density $\Omega h^2$.}
	\label{fig3}
\end{figure*}

In Fig. \ref{fig1}, we project the surviving samples in the light CP-odd Higgs mass $m_{a_1}$ versus LSP mass $m_{\tilde\chi^0_1}$ planes, with colors indicating DM relic density $\Omega h^2$, LSP singlino component $|N_{15}|^2$, LSP bino component $|N_{11}|^2$, $h_2$ invisible branching ratio $Br(h_2^{\rm inv})$, $h_1$ invisible branching ratio $Br(h_1^{\rm inv})$, and $h_1$ diphoton signal rate $R(h_1^{\gamma\gamma})$, respectively. 
From this figure, one can see that the LSP neutralino $\tilde\chi^0_1$ can be bino- or singlino-dominated, but there are also some samples with sizable higgsino component ($\gtrsim 0.1$). 
Both two types of $\tilde\chi^0_1$ can have four funnel annihilation mechanisms, i.e., annihilating through $Z$, $a_1$, $h_2$, and $h_1$. 
Different from these in our former work \cite{Wang:2020dtb}, $h_1$ in this work can be as heavy as $95\sim 100\GeV$ with $h_1$ funnel annihilation, and $\tilde\chi^0_1$ can be bino-dominated for non-universal gaugino masses. 
Also there are some samples with sufficient relic density, other than these with $m_{\tilde\chi^0_1}\lesssim 10\GeV$, or $Z, h_2$ funnels. 
The Higgs invisible branching ratios $Br(h_2^{\rm inv}), Br(h_1^{\rm inv})$ and the DM relic density $\Omega h^2$ are correlated, through the couplings between singlet-dominated $h_1, a_1$ and singlino-dominated $\tilde\chi^0_1$, which can be modified by adjusting the dimensionless parameters  $\lambda,\kappa$, or mixing with higgsino, as shown in Eq. (\ref{h1bobo}-\ref{a1soso}). 
Samples with sufficient relic density through the $h_i$ funnel annihilation mechanism usually get small invisible branching ratios of $h_i$. 

In Fig. \ref{fig2}, we project surviving samples in the $\kappa$ versus $\lambda$ planes, with the same color indications as these in Fig.\ref{fig1}. 
From Ref. \cite{Wang:2020tap}, one can know that for singlino-like $\tilde\chi^0_1$, singlet-like $h_1$ and $a_1$, and with small $\lambda,\kappa$ and not-too-large $m_{a_2}$,  there will be a correlation between these masses:  $m_{a_1}^2+m_{h_1}^2/3\approx m_{\tilde\chi^0_1}^2$.
But in this work we have the singlet-dominated $m_{h_1}\approx 95\GeV$, thus that can only be possible when $m_{\tilde\chi^0_1}\lesssim 54\GeV$. 
As can be seen from Fig. \ref{fig2}, when $\tilde\chi^0_1$ is singlino-dominated, there can be $|\kappa|\ll 0.1$ but $\lambda\gtrsim 0.3$. 
Thus the correlation is not compatible in this work. 
When $\tilde\chi^0_1$ is singlino-dominated, $m_{\tilde\chi^0_1}\approx 2|\kappa\mu_{\rm eff}/\lambda|$, with $\mu_{\rm eff} >100\GeV$ and $m_{\tilde\chi^0_1} \lesssim 64\GeV$ we have $|\kappa|/\lambda \lesssim 0.32$. 
From this figure, one can see that is true when the LSP $\tilde\chi^0_1$ is singlino-dominated, but not when it is bino-dominated. 

In Fig. \ref{fig3} we project surviving samples in the signal rates at future lepton colliders, $R(V h_2^{\rm inv})$, $R(V h_1^{\rm inv})$, and $R(V h_1^{\rm b\bar{b}})$, versus LSP mass $m_{\tilde\chi^0_1}$ planes, respectively, with colors indicating DM relic density $\Omega h^2$. 
From this figure one can clearly see that, samples with sufficient relic density are accompanied by low invisible signal rates $R(V h_i^{\rm inv})$ at future lepton colliders. 
But the $h_1$ at about $95\GeV$ can be checked through $Vb\bar{b}$ signal. 

\section{Conclusions}
\label{conc}

In light of the $95\GeV$ diphoton excess released by CMS, and DM direct searches by XENONnT and LZ collaboration last year, we study the correlation between light dark matter and a $95\GeV$ scalar. 
In this work, we study that in the GUTc-NMSSM, where the NMSSM has two doublets and one singlet Higgs superfield, and most parameters are input at the GUT scale, but with Higgs boson and gaugino masses not unified there, respectively. 
In the calculations, we also consider other recent experimental constraints, such as Higgs data, SUSY searches, DM relic density, etc. 
After detailed analysis and discussion, we find that: 
(i) The light DM can be bino- or singlino-dominated, but can be mixed with minor component of Higgsino. 
(ii) Both these two cases can get right relic density or sizable Higgs invisible decay, by adjusting the dimensionless parameters $\lambda, \kappa$, or suitably mixing with Higgsino. 
(iii) Both cases can have four funnel annihilation mechanisms, i.e., annihilating through $Z, a_1, h_2, h_1$, although with $h_1$ at about $95\GeV$. 
(iv) Samples with right relic density usually get a weak signal of Higgs invisible decay at future lepton colliders, but the  $95\GeV$ scalar can have sizable $b\bar{b}$ signal.

\section*{Acknowledgments.}
This work was supported by the National Natural Science Foundation of China (NNSFC) under Grant Nos. 12275066, 11605123, and 12074295.


\end{document}